\documentclass[12pt]{article}
\usepackage[T1]{fontenc}
\usepackage{lmodern}
\usepackage[english]{babel}

\usepackage{amsmath}
\usepackage{amssymb}
\usepackage{mathtools}
\usepackage{graphicx}

\usepackage{cite}
\usepackage{siunitx}
\usepackage{booktabs}
\usepackage{multirow}

\usepackage[labelfont=bf]{caption}
\usepackage[title]{appendix}

\usepackage{xfrac}
\usepackage[nodisplayskipstretch]{setspace}
\numberwithin{equation}{section}
\usepackage[activate={true,nocompatibility},final,
tracking=true,kerning=true,spacing=true,
factor=1100,stretch=10]{microtype}
\SetTracking{encoding={*}, shape=sc}{40}
\microtypecontext{spacing=nonfrench}

\usepackage[dvipsnames]{xcolor}
\definecolor{blue}{RGB}{0,0,128}
\definecolor{Azure}{HTML}{0F52BA}

\usepackage{hyperref}
\hypersetup{colorlinks,allcolors=Azure,pageanchor=false}
\usepackage[margin=1.05in]{geometry}%after hyperref
\usepackage{orcidlink}%after hyperref

\NewDocumentCommand{\email}{m}{\href{mailto:#1}{\texttt{#1}}}

\NewDocumentCommand{\muetoee}{}{\mu^-e^-\rightarrow{}e^-e^-}
\NewDocumentCommand{\mueee}{}{\mu\rightarrow{}3e}
\NewDocumentCommand{\mutoegam}{}{\mu\rightarrow{}e\gamma}
\NewDocumentCommand{\mutoea}{}{\mu\rightarrow{}ea}
\NewDocumentCommand{\mutoeinv}{}{\mu\rightarrow{}e +\mathrm{inv}}
\NewDocumentCommand{\muNtoeN}{}{\mu \mathrm{N}\rightarrow{}e\mathrm{N}}
\NewDocumentCommand{\MtoMbar}{}{M_{\mu}\rightarrow\overline{M_{\mu}}}
\NewDocumentCommand{\mutoegg}{}{\mu\rightarrow{}e\gamma\gamma}

\NewDocumentCommand{\order}{m}{\mathcal{O}{(#1)}}
\NewDocumentCommand{\inv}{}{~\mathrm{inv}}
\NewDocumentCommand{\br}{m}{\mathcal{B}{(#1)}}
\NewDocumentCommand{\gev}{}{\mathrm{GeV}}

\NewDocumentCommand{\gs}{}{g^S_{e\mu}}
\NewDocumentCommand{\gp}{}{g^P_{e\mu}}
\NewDocumentCommand{\gsp}{}{g^{S, P}_{e\mu}}
\NewDocumentCommand{\gpe}{}{g^P_{ee}}

\DeclarePairedDelimiter{\abs}{\lvert}{\rvert}

\begin{document}
\allowdisplaybreaks{}
\begin{titlepage}
  \begin{flushright} {\small
      USC-TH-2026-02\\
      \today}
  \end{flushright}%
  \vspace{9em}
  \begin{center}
    {\Large \bfseries
      Axionlike Particles and Lepton Flavor Violation\\[0.2em]
      in Muonic Atoms
    }\\[2.5em]
   { \scshape Girish Kumar\,\orcidlink{0000-0001-6051-2495} {\normalshape and}
     Alexey A.~Petrov\,\orcidlink{0000-0002-4945-4463}
   }\\[1.5em]
    {\itshape
      Department of Physics and Astronomy,\\ University of South Carolina,\\
      Columbia, South Carolina 29208, USA
    }\\[0.7em]
    {\textit{E-mail}: \email{girish89@sc.edu}, \email{apetrov@sc.edu} }
  \end{center}%
  \vspace{0.8em}
  %%%%%%%%%%%%%%%%%%%%%%%%%%%%%%%%%%%%%%%%%%%%%%%%%%%%%%%%%%%%%%%%%%%%%%%%%%%
  \begin{abstract}
    \noindent
    We explore the potential of the Mu2e experiment to probe the lepton-flavor-violating process $\mu^- e^- \to e^- e^-$ in a muonic atom within a simplified axionlike particle (ALP) framework featuring flavor-violating $e$-$\mu$ couplings and a flavor-diagonal pseudoscalar coupling to electrons, which also allows for possible invisible ALP decays into a dark sector. We compute the ALP-mediated contribution to the transition rate and show that, at fixed couplings, the branching ratio increases for lighter mediators and scales as $(Z-1)^3$, favoring heavier nuclei. We compare the model against constraints from $\mu\to e\gamma$, $\mu\to 3e$, $\mu\to e\gamma\gamma$, $\mu\to e+\mathrm{inv}$, and muonium-antimuonium conversion, as well as from the anomalous magnetic moments of the electron and muon. Additional astrophysical and beam-dump limits on the electron coupling are also discussed. A key result is that $\Delta a_e$ provides one of the most stringent probes of the parameter space and, in the global scan, excludes the largest fraction of sampled points. After applying the laboratory constraints used in the scan, the viable branching ratio for $\mu^- e^- \to e^- e^-$ in aluminum drops to at most $\mathcal{O}(10^{-20})$, while the resonant region $2m_e<m_a<m_\mu-m_e$ is much more heavily suppressed. The highest achievable values are closely tied to $\mathcal{B}(\mu\to 3e)$ near its current limit, indicating that the upcoming Mu3e experiment will explore the most promising region relevant for this muonic-atom signal. Our analysis shows that, although a light ALP can parametrically enhance $\mu^- e^- \to e^- e^-$ at fixed couplings, existing bounds---especially $\Delta a_e$, $\mu\to 3e$, $\mu\to e\gamma$, and muonium oscillations---severely limit the observable rate.
  \end{abstract}
  %%%%%%%%%%%%%%%%%%%%%%%%%%%%%%%%%%%%%%%%%%%%%%%%%%%%%%%%%%%%%%%%%%%%%%%%%%%%
  % \Textbf{ Keywords:} Lepton flavor violation, light new physics, axionlike
  % particles, dark sector
\end{titlepage}

%%%%%%%%%%%%%%%%%%%%%%%%%%%%%%%%%%%%%%%%%%%%%%%%%%%%%%%%%%%%%%%%%%%%%%%%%%%
\section{Introduction}\label{sec:intro}
%%%%%%%%%%%%%%%%%%%%%%%%%%%%%%%%%%%%%%%%%%%%%%%%%%%%%%%%%%%%%%%%%%%%%%%%%%%

Particle interactions in the Standard Model (SM) conserve individual lepton flavors, but observations of neutrino oscillations (see Ref.~\cite{Gonzalez-Garcia:2002bkq} for a review) show that lepton flavor is broken in nature. This provides one of the strongest reasons for physics beyond the SM. Including finite neutrino masses and mixing, however, does not necessarily cause flavor violation in the charged lepton sector at an observable level \cite{Echenard:2026cdv}. For example, the SM charged current interaction involving Dirac neutrinos can induce the lepton-flavor-violating (LFV) process $\mutoegam$ at one-loop level, but its decay rate is proportional to the neutrino mass-squared difference, a result of the Glashow–Iliopoulos–Maiani (GIM)-mechanism~\cite{Glashow:1970gm}, making its value smaller than $\order{10^{-54}}$~\cite{Petcov:1976ff,Marciano:1977wx}, far below current and future experimental sensitivity. However, this extreme suppression means experimental searches for charged LFV processes are practically free of SM background, so any detection would indicate new physics. Interestingly, many well-motivated new physics (NP) models include interactions that do not experience GIM-like suppression and can produce LFV processes at rates accessible to experiments.

The strongest bounds on charged lepton flavor violation come from muon LFV processes\footnote{In this paper, we focus on muon LFV processes. For a more complete list of LFV processes with their recent experimental status, see, for example, Ref.~\cite{Frau:2024rzt}.}. The leading limit is from the decay $\mutoegam$; the MEG~II experiment~\cite{MEGII:2025gzr} recently published results for $\mutoegam$ search and set an upper limit on its branching ratio, $ \br{\mutoegam} < 1.5 \times 10^{-13}$, about a factor of 2 improvement over the previous limit~\cite{MEGII:2023ltw,MEG:2016leq}. Meanwhile, the current upper bound of $10^{-12}$ on the branching ratio of the three-body muon decay $\mueee$ is from an older measurement by the SINDRUM experiment~\cite{SINDRUM:1987nra}. Another significant LFV process is muon-to-electron conversion in nuclei $\muNtoeN$, where a muon captured in the nucleus converts into an electron without neutrino emission; the best limit on this process, established by the SINDRUM II experiment~\cite{SINDRUMII:2006dvw} using gold as the target, is $R_{\mu e} < 7 \times 10^{-13}$, with $R_{\mu e}$ representing the muon-to-electron conversion rate normalized to the total muon capture rate.

While the above processes are examples of lepton flavor violation by one unit, many NP models feature interactions that violate lepton flavor by more than one unit, leading to exotic LFV channels~\cite{Heeck:2024uiz,Heeck:2025jfs}. Muonium to antimuonium transition $\MtoMbar$ is an important example of such processes that violate both electron and muon flavor by two units; the current best limit on the probability of $\MtoMbar$ transition is $\order{10^{-10}}$ by the MACS experiment at PSI \cite{Willmann:1998gd}.

If NP particles with LFV couplings are light, they can also be produced in the final state. For example, the semi-invisible muon decay $\mu\rightarrow e\inv$, where ``inv'' denotes one or multiple invisible particles, is a unique signature in NP models involving light, invisible NP particles such as axionlike particles. For 2-body decay, the current bound is $\br{\mutoeinv}\lesssim\order{10^{-5}}$ \cite{Jodidio:1986mz,TWIST:2014ymv}. However, if the light NP particle has a shorter lifetime, so that it decays within the detector to visible states, for example, $ee$, $\gamma\gamma$, then, as reviewed later in the text, 3-body muon decays $\mueee$ and $\mutoegg$ give superior limits, especially if the decay is resonant.

In terms of improvements in future sensitivity, the most promising channels seem to be $\mueee$ and $\muNtoeN$. The Mu3e experiment at PSI \cite{Blondel:2013ia} will enhance the current limit on $\mueee$ by four orders of magnitude to $10^{-16}$. Two experiments, Mu2e at Fermilab \cite{Mu2e:2014fns} and COMET at J-PARC \cite{COMET:2018auw}, aim to measure $\muNtoeN$ with a maximum sensitivity of $10^{-17}$, while another experiment, DeeMe~\cite{Teshima:2019orf}, will measure it at a level of $10^{-14}$. In Table~\ref{tab:lfv}, we provide a summary of the current best limits and projected sensitivities of these channels in the near future.

%%%%%%%%%%%%%%%%%%%%%%%%%%%%%%%%%%%%%%%%%%%%%%%%%%%%%%%%%%%%%%%%%%%%%%%%%%%
{\renewcommand{\arraystretch}{1.3}%
  \begin{table}
    \centering
    \caption{Summary of the current experimental upper limits ($90\%$ CL)
      on branching ratios of muon flavor-violating processes and their projected
      future sensitivities. Note that bounds on $\mutoeinv$ are for a two-body
      decay, with ``$\text{inv}$'' denoting a neutral scalar boson particle
      such as ALPs, and depend on the chirality of the ALP-lepton
      interaction~(see Ref.~\cite{Calibbi:2020jvd}). Also, note that the bound
      $\MtoMbar$ is for the muonium to antimuonium conversion probability,
      which depends on the magnetic field in the experimental apparatus (0.1
      Tesla for the MACS experiment~\cite{Willmann:1998gd}). The magnetic field
      dependence of different interaction types is accounted for by correction
      factors $S_{B}$ which are given in Table II of
      Ref.~\cite{Willmann:1998gd}).}
    \label{tab:lfv}
    \begin{tabular}{@{} l @{\hspace{2em}} c c @{\hspace{2em}} c c @{}}
      \toprule
      \multirow{2}[3]{*}{Process}
      & \multicolumn{2}{c}{Current best limit}  
      & \multicolumn{2}{c}{Future sensitivity}\\
      \cmidrule(l{1em}r{2em}){2-3}\cmidrule(lr){4-5}
      &\multicolumn{1}{c}{Upper limit} &{Experiment}
      &\multicolumn{1}{c}{Value} & {Experiment} \\
      \midrule
      $\mutoegam$
      &  $1.5 \times 10^{-13}$ & MEG~II~\cite{MEGII:2025gzr}
      & $6\times 10^{-14}$     & MEG~II~\cite{MEGII:2025gzr} \\
      $\mueee$
      & $1.0\times 10^{-12}$   & SINDRUM~\cite{SINDRUM:1987nra}
      & $10^{-16}$             & Mu3e~\cite{Blondel:2013ia}\\
      $\muNtoeN$
      &  $7 \times 10^{-13}$   & SINDRUM~II~\cite{SINDRUMII:2006dvw}
      & $10^{-17}$             & Mu2e~\cite{Mu2e:2014fns} \\
      & & & $10^{-17}$         & COMET~\cite{COMET:2018auw}\\
      % & & && $10^{-14}$
      % & DeeMe~\cite{Teshima:2019orf}\\
    $\mutoeinv$
    & $\sim 10^{-5}$          & TWIST~\cite{TWIST:2014ymv}
    & $\sim  10^{-8}$         & Mu3e~\cite{Perrevoort:2018okj}\\
    & $\sim 10^{-6}$          & Jodidio~\textit{et al.}~\cite{Jodidio:1986mz}
    & &\\
    $\mutoegg$
    & $7.2 \times 10^{-11}$   & Crystal Box~\cite{Bolton:1988af}
    & \dots  & \dots\\
    $\MtoMbar$
    & $8.3 \times 10^{-11}$  & MACS~\cite{Willmann:1998gd}
    & $3.8 \times 10^{-13}$  & MACE~\cite{Bai:2024skk}\\
    \bottomrule
  \end{tabular}%
\end{table}}%
%%%%%%%%%%%%%%%%%%%%%%%%%%%%%%%%%%%%%%%%%%%%%%%%%%%%%%%%%%%%%%%%%%%%%%%%%%%

If NP particles mediating LFV processes are much heavier than the intrinsic energy scale of the process, e.g., $\sim \order{1}~\gev$ for muon processes, NP effects can be described using LFV local operators within an effective field theory framework. Processes like $\mutoegam$, $\mueee$, and $\muNtoeN$ would be the best candidates for searching for $\mu-e$ flavor violation. Remember that $\mutoegam$ is only sensitive to dipole operators, $\mueee$ to four-lepton contact operators, and $\muNtoeN$ to $2q2\ell$ local operators (the latter two are also sensitive to $\mu-e-\gamma^{\ast}$ dipole operators). The muonium-antimuonium transition is, on the other hand, most suitable for probing $\Delta L=2$ operators \cite{Conlin:2020veq,Conlin:2022sga,Petrov:2022wau}. However, as mentioned earlier, for light NP particles, LFV channels such as $\mu\rightarrow{}ea$ would be the most sensitive. Therefore, it is clear that understanding the underlying NP requires studying different LFV channels.

With this in mind, in this article, we aim to focus on a rather exotic and nuclei-assisted LFV process, $\muetoee$ in a muonic atom \cite{Koike:2010xr}. In experiments such as muon-to-electron conversion, muons stop in the target material, with the captured muon cascading down to the $1s$ level of an $Al$ atom, replacing one of the electrons to form a muonic atom. Then, there is a small but finite probability that the bound muon can scatter with the other $1s$ electron. Kinematically, only a few final states are accessible in this scattering, as the total center-of-mass energy is close to the muon's rest mass. In the presence of the same LFV NP interactions that induce $\mueee$, the initial bound muon and electron can scatter into two electrons. The experimental signature is clear since it involves charged particles only and no photons, and the phase space is larger compared to $\mueee$. Experimentally, the search for this process can be pursued within the stopped-muon environments of ongoing and forthcoming $\muNtoeN$ programs. In particular, there will be an opportunity to search for this process at the COMET and  Mu2e experiment; a brief discussion about the possibility of carrying out $\muetoee$ measurement in phase I of the COMET experiment can be found in Ref.~\cite{COMET:2018auw}, and similar discussions are also currently underway at the Mu2e experiment.

This process has been previously examined within the framework of effective field theory relevant for studying heavy NP effects \cite{Koike:2010xr,Uesaka:2016vfy,Uesaka:2017yin,Uesaka:2018emk,Kuno:2019ttl} \footnote{The process has also been discussed in a few specific NP models such as the SM extension by heavy sterile neutrino~\cite{Abada:2015oba}, and models based on hidden $U(1)$ symmetry \cite{Nomura:2020azp,Nomura:2020dzw}.}. The process $\muetoee$ receives contributions from four-lepton contact operators and photonic-dipole operators. Since these operators are strongly constrained by LFV decays $\mueee$ and $\mutoegam$, respectively, it was found~\cite{Koike:2010xr} that the allowed upper bound on the branching ratio is very small, $\sim \order{10^{-20}}$ for aluminum. However, if the NP particles mediating this process are light, the predicted branching ratio could be enhanced because the suppression due to a heavy NP mass scale is lifted. The purpose of our study is to extend previous analyses of this process, performed under the assumption of heavy NP, to the case of light NP mediators. Axionlike particles (ALPs) are one example of well-motivated light NP particles, which have recently received significant attention in the literature, especially regarding charged lepton flavor violation~\cite{Heeck:2017xmg,Bauer:2019gfk,Cornella:2019uxs,Calibbi:2020jvd,Endo:2020mev,Cheung:2021mol,Bauer:2021mvw,Cui:2021dkr,Haghighat:2021djz,Davoudiasl:2021haa,Knapen:2023zgi,Knapen:2024fvh,Batell:2024cdl,Calibbi:2024rcm,Davoudiasl:2024vje,Fuyuto:2024skf,Ema:2025bww,MartinCamalich:2025srw}. In this paper, we investigate the contribution of light ALPs to this process $\muetoee$ and examine the allowed upper limit on this process consistent with current data.

The rest of the article is organized as follows. In the next section, we introduce a minimal ALP model with LFV couplings relevant for $\muetoee$ and set up our notations. In section \ref{sec:br}, we describe the ALP contribution to $\muetoee$ and derive the expression for its rate. In section \ref{sec:constraints}, we discuss observables that constrain the parameter space of the model. Our main results are presented in section \ref{sec:results}, where we analyze the allowed values of the $\muetoee$ branching ratio in the model and show correlations with other observables. Finally, our conclusions are provided in section~\ref{sec:conclusions}.

%%%%%%%%%%%%%%%%%%%%%%%%%%%%%%%%%%%%%%%%%%%%%%%%%%%%%%%%%%%%%%%%%%%%%%%%%%%
\section{The Model}\label{sec:model}%
%%%%%%%%%%%%%%%%%%%%%%%%%%%%%%%%%%%%%%%%%%%%%%%%%%%%%%%%%%%%%%%%%%%%%%%%%%%
At low energies and up to the operators of dimension-5, general couplings between an ALP field $a$ and charged leptons $\,\ell_{i}$ in the mass basis are described by the following effective Lagrangian \cite{Georgi:1986df,Cornella:2019uxs}
\begin{equation}
  \label{eq:general-Lag}
  \mathcal{L}_{eff}
  =
  -\frac{\partial_{\mu}a}{\Lambda}
  \sum_{i, j} \bar \ell_i\gamma^{\mu}
  \left( v^{\ell}_{ij} - a^{\ell}_{ij} \gamma_5\right)
  \ell_{j}\,,
\end{equation}
where $v^{\ell}_{ij}$ and $a^{\ell}_{ij}$ denote vector and axial-vector couplings, respectively. The corresponding $3\times 3$ matrices in flavor space, $v^{\ell}$ and $a^{\ell}$, are Hermitian and real, which implies that $v^{\ell}_{ij}=v^{\ell}_{ji}$ and $a^{\ell}_{ij}=a^{\ell}_{ji}$. As already implicit from the derivative nature of interactions in the Lagrangian above, ALPs are pseudo-Nambu-Goldstone boson particles, remnants of the spontaneous breaking of a global symmetry at a high scale. Therefore, the mass of an ALP, $m_{a}$, is naturally small compared to the broken scale $\Lambda$ in Eq.~\eqref{eq:general-Lag}.

% ALP generally couples to quarks and gauge bosons as well, which means the number of independent parameters in the model is very large. Since we are interested in the process $\muetoee$, to simplify the phenomenological analysis, we will focus only on a subset of ALP-lepton interactions—specifically, those that contribute to this process at the leading order. Therefore, we work with a simplified model in which the ALP couples only to the lightest leptons, described by the following Lagrangian,
In general, ALP couplings to leptons, quarks, and gauge bosons may be
present. Allowing for all such interactions introduces a large number of
independent parameters. Many of these interactions do not contribute directly
to the process considered here, although they may lead to additional
experimental signatures and constraints.
 Since we are interested in the
process $\muetoee$, to simplify the phenomenological analysis, we focus only
on the subset of ALP-lepton interactions that contributes to this process at
leading order. Therefore, we work with a simplified model in which the ALP
couples only to the lightest leptons, described by the following Lagrangian,

\begin{align}
  \label{eq:Lag-alp}
  \mathcal{L}_{\text{eff}}
  =
  \left\{
  -i a \bar{e} \left( \gs + \gp\gamma_5 \right)\mu
  +\text{H.c.}
  \right\}
  - i g^{P}_{ee}\, a \bar e \gamma_5 e\,,
\end{align}
which contains LFV couplings in $e$-$\mu$ sector and a flavor-conserving
coupling to electrons. Here, for convenience and brevity of notations we have
introduced dimensionless scalar $ g^{S}_{\ell_i\ell_j}$ and pseudoscalar
couplings $ g^P_{\ell_i\ell_j}$.  One can show that this Lagrangian is
equivalent to the one in Eq.~\eqref{eq:general-Lag} after applying Dirac
equation on lepton fields, up to a shift in the ALP coupling to photons arising
due to divergence of axial-vector current. Also note that vector current
conservation implies that flavor-diagonal couplings $v^{\ell}_{ii}$ (or
$g^{S}_{\ell_i \ell_{j}}$) do not exist. Couplings
$ g^{S, P}_{\ell_i\ell_j}$ are related to the vector and axial-vector
couplings in the Lagrangian in Eq.~\eqref{eq:general-Lag} via
\begin{align}
\label{eq:coupling-relations}
  g^S_{\ell_i\ell_j}
  = v^{\ell}_{ij}\frac{m_{\ell_j}-m_{\ell_i}}{\Lambda}\,,
  \qquad
  g^P_{\ell_i\ell_j}
  = a^{\ell}_{ij}\frac{m_{\ell_j}+m_{\ell_i}}{\Lambda}.
\end{align}%

Flavor-diagonal ALP couplings to SM fermions induce $a\to 2\gamma$ at one loop~\cite{Bauer:2017ris}, modifying the ALP-photon coupling $c_{\gamma\gamma}$ in the UV theory. We assume $c_{\gamma\gamma}$ arises solely from this loop contribution. Describing the ALP-photon coupling via the interaction $c^{\text{loop}}_{\gamma\gamma} (\alpha/2\pi)\, a\, F_{\mu\nu}\tilde{F}^{\mu\nu}$, the finite one-loop contribution from the electron coupling $\gpe$ is
\begin{equation}
  \label{eq:c-gamgam}
  c^{\text{loop}}_{\gamma\gamma}
  =
  \frac{1}{2m_e} \gpe\, B_1(\tau_e)\,,
\end{equation}%
where $\tau_e = 4 m^2_{e}/m^2_a - i \epsilon$, and the loop function
$B_1(\tau_e)$ is defined in the Appendix~\ref{app:loop}.

With ALP interactions as defined above, depending on the ALP mass regime, the
total width of an ALP is given as
\begin{equation}
\label{eq:alp-width}
  \Gamma_a
  =
  \begin{dcases}
    \Gamma(a\rightarrow{}\gamma\gamma)
    +  \Gamma(a\rightarrow{}ee)
    +  \Gamma(a\rightarrow{}e\mu)
    & \text{for } m_a> m_{\mu} + m_e\,,\\
    \Gamma(a\rightarrow{}\gamma\gamma)
    +  \Gamma(a\rightarrow{}ee)
    & \text{for } 2 m_e < m_a < m_{\mu} + m_e\,,\\
    \Gamma(a\rightarrow{}\gamma\gamma)
    & \text{for } m_a < 2 m_e\,.
  \end{dcases}
\end{equation}

ALPs can also interact with new particles beyond the SM, leading to a non-negligible width not generated by their decays into SM particles. The simplest way to account for this is to assume that the ALP couples to a dark-sector (DS) fermion $\chi$ with mass $m_\chi$. Their interaction is described by the Lagrangian
\begin{align}
  \label{eq:ALP-DM}
  \mathcal{L}_{a\chi\chi} =   - i g_{\chi\chi} a \,\bar \chi \gamma_5 \chi\,,
\end{align}
where, for simplicity, $\chi$ is taken to be a Dirac fermion. Although the coupling $g_{\chi\chi}$ does not directly contribute to $\muetoee$, this interaction opens a new ALP decay channel, $a\to \chi\bar \chi$, when $m_a > 2 m_\chi$. This contributes to the total ALP width in Eq.~\eqref{eq:alp-width} and  can modify the interplay among the LFV constraints. We will use it as a minimal benchmark to test whether an invisible ALP width can qualitatively change our conclusions.  Note that unless specified otherwise, $g_{\chi\chi}$ is always assumed to vanish in our analysis.

%%%%%%%%%%%%%%%%%%%%%%%%%%%%%%%%%%%%%%%%%%%%%%%%%%%%%%%%%%%%%%%%%%%%%%%%%%%
\section{ALP-mediated \texorpdfstring{$\boldsymbol{\muetoee}$}{mue->ee} in
  Muonic Atom}\label{sec:br}%
%%%%%%%%%%%%%%%%%%%%%%%%%%%%%%%%%%%%%%%%%%%%%%%%%%%%%%%%%%%%%%%%%%%%%%%%%%%

\begin{figure}[b]
  \centering
  \includegraphics[scale=0.12]{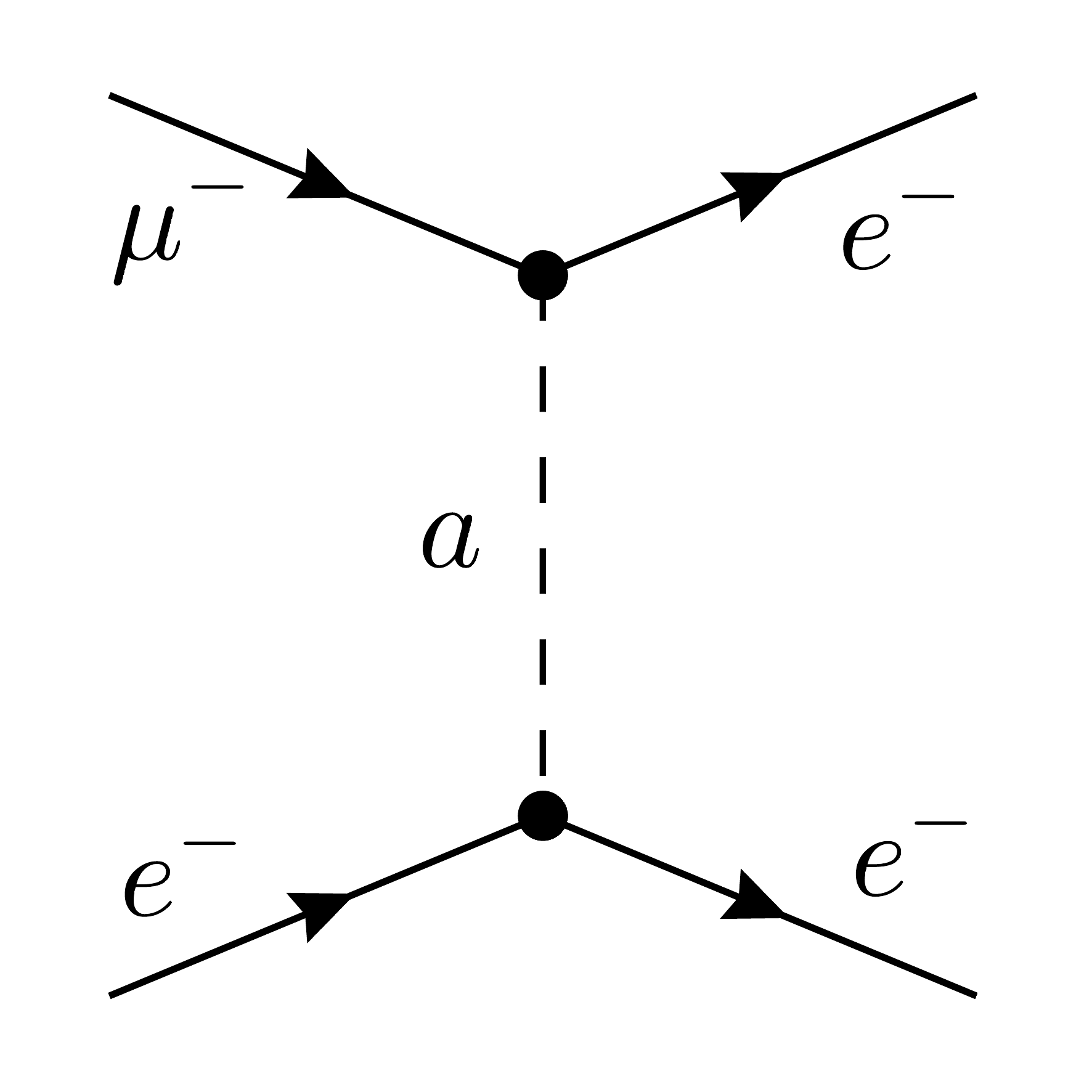}\qquad
  \includegraphics[scale=0.12]{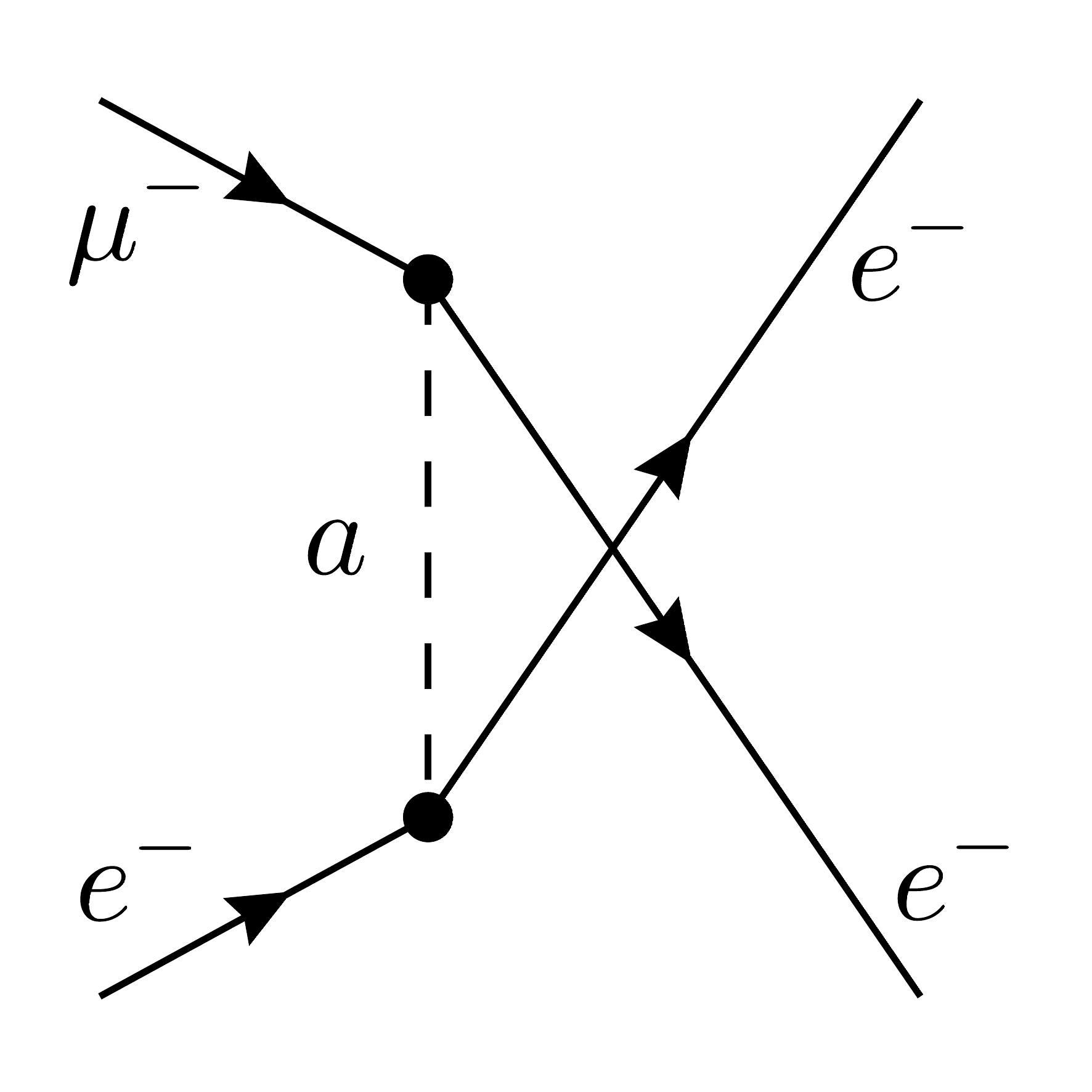}
  \caption{ALP-induced $\muetoee$ at tree-level.}%
  \label{fig:feyn_mue2ee}
\end{figure}

The lowest order interaction between the muon and atomic electrons can be seen
in Fig.~(\ref{fig:feyn_mue2ee}), where the ALP propagator in momentum space,
including a finite width $\Gamma_a$, is taken as
\begin{equation}
\Delta(q)
=
\frac{i}{q^2 - m_a^2 + i m_a \Gamma_a}.
\label{eq:propagator}
\end{equation}
with $\Gamma_a$ modeled following Eq.~\eqref{eq:alp-width}.

We work in momentum space and follow the approach used in Ref.~\cite{Koike:2010xr} to calculate the transition rate of $\muetoee$ in a muonic atom. The process receives contributions from $t$- and $u$-channel ALP exchange, and the total amplitude $\mathcal{M}$ is equal to their difference since the final state involves identical electrons. Ignoring for a moment the Coulomb field of nuclei, the cross-section for $\muetoee$ is calculated using the standard $2\rightarrow 2$ differential cross-section formula
%%%%
\begin{align}
   \label{eq:def-cross-section}
  d\sigma\, \abs{\mathbf{v}_1 - \mathbf{v}_2}
  =
  \frac{1}{(2E_1)(2E_2)} \frac{1}{2}
  d\Pi_\text{LIPS} \,\abs*{\overline{\mathcal{M}}}^2\,,
\end{align}
%%%%%
where $E_{1, 2}$ are the energies of initial particles and $d\Pi_\text{LIPS}$
denotes the Lorentz invariant phase space. The spin-averaged squared amplitude,
$\abs*{\overline{\mathcal{M}}}^2$, can be evaluated using
\texttt{FEYNCALC}~\cite{Shtabovenko:2023idz}. After integrating over the phase space, we
obtain the following cross section for $\muetoee$:
\begin{align}
  \label{eq:cross-section}
  \sigma\, \abs{\mathbf{v}_1 - \mathbf{v}_2}
  &=
    \frac{
    (m_{\mu}-m_e)
    \sqrt{{(m_\mu+m_e)}^2- 4m_e^2}
    }{
    64\pi(m_\mu+m_e)
    \left[
    {(m_e(m_\mu-m_e)+ m_a^2)}^2 + m_a^2\Gamma_a^2
    \right]}\notag\\
  &\quad\quad \times
    \abs{\gpe}^2\left\{
    m_\mu (\abs{\gp}^2 + 3 \abs{\gs}^2)
    - m_e (\abs{\gp}^2 -9 \abs{\gs}^2)
    \right\},\notag\\
  &\simeq
    \frac{m^2_{\mu}\  \abs{\gpe}^2
    \left(\abs{\gp}^2 + 3 \abs{\gs}^2\right)
    }{64\pi\left[
    {(m_em_\mu+ m_a^2)}^2 + m_a^2\Gamma_a^2
    \right]}\,,
\end{align}
where in the writing the second line, terms of $\order{m_e/m_{\mu}}$ have been
ignored. In order to derive the above expression, we have ignored the 3-momenta
of bound muon and atomic electron. The final electrons propagate back-to-back
and each has energy equal to about half of muon mass.

The decay rate for $\muetoee$ is then calculated by multiplying the cross-section $\sigma \abs{\mathbf{v}_1 - \mathbf{v}_2}$ with the probability of having an initial bound muon and an atomic electron at the same position~\cite{Koike:2010xr,Uesaka:2018emk}. Denoting the bound muonic and electronic wave functions of the $1s$ state by $\psi_{\mu}( r)$ and $\psi_e( r)$, respectively, the probability is given by $\int \!\mathrm{d}^3 r\, \abs{\psi_{\mu}( r)}^2 \abs{\psi_e( r)}^2$.  Since the bound muon, in comparison to the atomic electron, is much ``closer'' to the nucleus (due to the heavier muon mass), the muon density $ \abs{\psi_{\mu}( r)}^2$ is approximated to a Dirac delta function $\delta^3( r)$.  With this approximation, and considering that the $1s$ state can accommodate two electrons, and that the nuclear charge $Z$ is screened by the muon, the $\muetoee$ decay rate is obtained as~\cite{Koike:2010xr}
\begin{align}
  \label{eq:decay-rate}
  \Gamma(\muetoee)
  =
  2\, \abs{\psi_e(0)}^2\,\sigma\,
  \abs{\mathbf{v}_1 - \mathbf{v}_2}\,,
\end{align}
with the nonrelativistic wave function for the $1s$ electron given by
\begin{equation}
  \label{eq:1s}
\psi_e(r)
=
\frac{1}{\sqrt{\pi a_e^3}}\,e^{-r/a_e},
\qquad \text{with }
a_e = \frac{1}{(Z-1)\alpha\, m_e}\,.
\end{equation}

\begin{figure}
  \centering
  \includegraphics[width=0.7\textwidth]{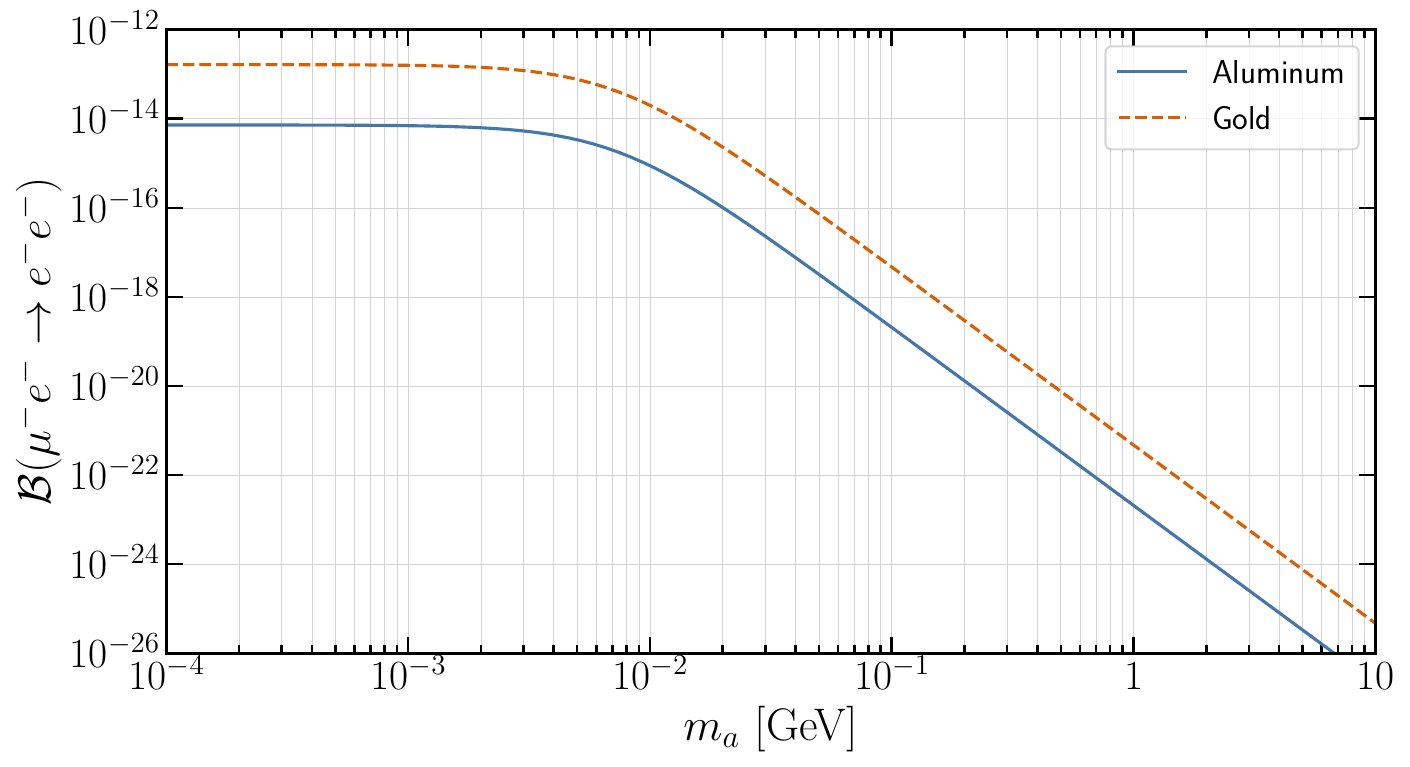}
  \caption{ALP mass dependence of branching ratio of $\muetoee$ in muonic atom
    for aluminum and gold target, taking an example benchmark value
    $\gsp=\gpe=10^{-6}$.}
  \label{fig:br-m}
\end{figure}

For studying experimental sensitivity, it is more convenient to discuss results
in terms of branching ratio of this process, obtained by dividing the decay rate
of $\muetoee$ by the total decay rate of muonic atom $1/\tilde\tau_{\mu}$,
where $\tilde \tau_\mu$ is the lifetime of muonic atom. The value of
$\tilde \tau_{\mu}$ for various elements are tabulated in
Ref.~\cite{Suzuki:1987jf}; for example, for hydrogen, aluminum, and gold,
$\tilde \tau_\mu = 2.19 \times 10^{-6}~\mathrm{s}$,
$ \sim 9 \times 10^{-7}~\mathrm{s}$, and $ \sim 7 \times 10^{-8}~\mathrm{s}$,
respectively.  Using Eqs.~\eqref{eq:cross-section}--\eqref{eq:1s}, the branching
ratio for $\mu^-e^-\to e^-e^-$ is then given by
\begin{align}
\label{eq:br-muonic}
  \mathcal{B}{(\mu^-e^-\to e^-e^-)}
   &=  \tilde \tau_\mu \, \Gamma(\muetoee)
   = \tilde \tau_\mu \, 2\, \abs{\psi_e(0)}^2
      \sigma \abs{\mathbf{v}_1 - \mathbf{v}_2}\,,\notag\\
   & \simeq
     \tilde \tau_\mu (Z-1)^3
     \frac{\alpha^3 m^3_e m^2_{\mu} \,
     \abs{\gpe}^2
     \left(
     \abs{\gp}^2 + 3 \abs{\gs}^2
     \right)}{32\pi^2
     [{(m_em_\mu+ m_a^2)}^2 + m_a^2\Gamma_a^2]}.
\end{align}
As can be noted from the above expression, and also valid
model-independently~\cite{Koike:2010xr}, the rate of $\muetoee$ is enhanced by a
factor of $(Z-1)^3$, which also underlines the advantage of searching this
process with muonic atoms (specially in heavy elements) in contrast to, for
example, a similar process with the muonium. In Fig.~\ref{fig:br-m}, taking
the benchmark values $\gsp=\gpe=10^{-6}$, we show the predicted branching ratio
$\mathcal{B}{(\mu^-e^-\to e^-e^-)}$ as a function of $m_a$ for aluminum
($Z=13$) and gold ($Z=79$) targets. The branching ratio is larger for the heavier
target due to the $(Z-1)^3$ enhancement factor discussed above. As expected, the
figure also shows that lighter ALPs enhance the $\mu^-e^-\to e^-e^-$ rate, while
the rate becomes increasingly suppressed as $m_a$ increases.

The results for the cross section and rate of $(\mu^-e^-\to e^-e^-)$ are derived under several simplifying assumptions. Following Ref.~\cite{Koike:2010xr}, we use nonrelativistic wave functions for bound leptons and treat scattered electrons as plane waves. Coulomb interactions are neglected, and only scattering with $1s$ electrons is included. These approximations hold well for sufficiently small $Z$, such as aluminum, where $Z\alpha \ll 1$. An improved analysis in Refs.~\cite{Uesaka:2016vfy,Uesaka:2017yin}, which employed relativistic wave functions for bound leptons and included Coulomb effects on both bound and scattered states, found that contact interactions enhance the rate more rapidly than $(Z-1)^3$, while photonic contributions show the opposite behavior. Overall, the improved $(\mu^-e^-\to e^-e^-)$ rate for light nuclei with $Z<20$, as expected, remains similar, although results for very heavy nuclei can differ by up to about an order of magnitude. Furthermore, including contributions from $2s$, $3s$,~$\dots ns$ electrons would affect the total rate by $\sim 20\%$~\cite{Koike:2010xr,Uesaka:2016vfy}. In summary, our results for $(\mu^-e^-\to e^-e^-)$ for light nuclear targets provide a reasonably accurate estimate.

In Fig.~\ref{fig:br-m}, we use a fixed ALP coupling, independent of the ALP mass. However, strong bounds on the ALP couplings from $\mu\to e$ LFV processes, discussed in the Introduction, must be taken into account to determine the allowed values of $\mathcal{B}{(\mu^-e^-\to e^-e^-)}$. These bounds are particularly stringent for light ALPs. In the next section, we review the relevant constraints on our model.

%%%%%%%%%%%%%%%%%%%%%%%%%%%%%%%%%%%%%%%%%%%%%%%%%%%%%%%%%%%%%%%%%%%%%%%%%%%
\section{Constraints}\label{sec:constraints}%
%%%%%%%%%%%%%%%%%%%%%%%%%%%%%%%%%%%%%%%%%%%%%%%%%%%%%%%%%%%%%%%%%%%%%%%%%%%

\textbf{\textit{LFV muon decays.}} The couplings $\gsp$, together with $\gpe$,
generate $\mutoegam$ via a one-loop triangle diagram involving the ALP and electron.
A similar loop involving the muon is not possible in our setup. The loop-induced
effective ALP-to-photon coupling in Eq.~\eqref{eq:c-gamgam} also contributes to
$\mutoegam$ at one loop and provides the dominant contribution. Ignoring terms of
$\order{m_e^2/m^2_\mu}$, the total $\mutoegam$ decay rate is given by
\begin{align}
  \label{eq:br-mu2egam}
  \Gamma(\mutoegam)
  =
  \frac{\alpha\, m_{\mu}}{2048\pi^{4}}\,
  \abs{\gpe}^{2}
  \left(
  \abs{\gs}^{2} \abs{\Delta_+}^2
  + \abs{\gp}^{2} \abs{\Delta_-}^2
  \right),
\end{align}
with $ \Delta_{\pm} = g_2(x) \pm (\alpha/\pi)(m_{\mu}/m_e) B_1(\tau_e) g_\gamma(x)$,
where $x=m^2_a/m^2_{\mu} -i\epsilon$,  and the loop functions $g_2(x)$, $g_\gamma(x)$, and
$B_1(\tau_e)$ are given in Appendix~\ref{app:loop}. Note that $g_\gamma(x)$ contains a
term depending logarithmically on the cutoff scale $\Lambda$, which we take to be
$\Lambda = 100~\gev$ in the numerical analysis.

For $m_a < m_{\mu}-m_e$, the couplings $\gsp$ enable on-shell ALP production in muon
decay, $\mutoea$, with the following decay rate:
\begin{align}
  \label{eq:br-mutoea}
  \Gamma(\mutoea)
  =
  \frac{m_{\mu}}{16\pi}
  \left(1-\frac{m^2_a}{m^2_{\mu}}\right)^2
  \left(\abs{\gp}^2 + \abs{\gs}^2\right)\,
\end{align}
The ALP can subsequently decay into $ee$ and $\gamma\gamma$ final states, resulting in resonant three-body decay processes $\mueee$ and $\mutoegg$. Working under the narrow-width approximation, the decay rate of $\mu\to eee$ via the resonant decays $\mutoea \to eee$, is given by
\begin{align}
  \label{eq:br-mu3e-on}
  \Gamma(\mueee)
  =
  \Gamma(\mu\rightarrow{}ea)
  \frac{\Gamma(a\rightarrow{}ee)}{\Gamma_a}\,,
\end{align}
valid in the ALP mass region $2m_e < m_a < m_{\mu} - m_e$, with the ALP decay
width into electrons given by
\begin{align}
  \label{eq:br-a2ee}
  \Gamma(a\rightarrow{}e^+e^-)
  =
  \frac{m_a}{8\pi}
  \abs{\gpe}^{2}
  \sqrt{1-\frac{4m^2_e}{m^2_a}}\,.
\end{align}

In calculating the $\mueee$ decay rate in the on-shell ALP region, we must
account for the finite ALP lifetime, as some produced ALPs may decay outside the
detector and go undetected. To this end, we multiply the decay rate in
Eq.~\eqref{eq:br-mu3e-off} by a factor $1-\mathcal{P}(d)$, where
$\mathcal{P}(d) \equiv e^{-d/\gamma c \tau}$ is the probability that the ALP does
not decay before traveling a distance $d$. The boosted decay length $\gamma c \tau$
of the ALP is calculated using the following expressions~\cite{Heeck:2017xmg,Cornella:2019uxs}:
\begin{align}
  \label{eq:alp-decay-length}
  \gamma c \tau = \frac{c\abs{\mathbf{p}_a}}{m_a \Gamma_a},\qquad
  \text{with}\enspace{}
  \mathbf{p}_{a} = \frac{\sqrt{\lambda(m_a^2, m_{\mu}^2, m_e^2)}}{2m_{\mu}}.
\end{align}
Here, $\mathbf{p}_a$ is the ALP three-momentum in the lab frame (the expression
above assumes the parent muon decays at rest), and $\lambda(a, b, c)$ denotes the
K\"all\'en function. We take $d \approx 1~\mathrm{m}$ as the typical detector size
in a $\mu\to 3e$ experiment.

For an on-shell ALP, Eq.~\eqref{eq:br-mu3e-on} provides the dominant contribution
to the $\mu\to 3e$ partial width. Another contribution arises from the
$\mu e\gamma^{\ast}$ dipole, which we calculate using the well-known
model-independent formula~\cite{Kuno:1999jp}:
\begin{align}
  \label{eq:dipole-mu3e}
  \frac{\Gamma(\mu\rightarrow{}e\gamma^{\ast}\rightarrow{}3e)}{\Gamma(\mutoegam)}
  \simeq
  \frac{\alpha}{3\pi}
  \left(
  \log \frac{m_{\mu}^2}{m_e^2}
  -\frac{11}{4}
  \right).
\end{align}
This contribution is present for both light and heavy ALPs, and becomes important
below the $2m_e$ threshold, where ALPs can no longer decay promptly to electrons.

Above the on-shell ALP threshold, $m_a > m_{\mu}-m_e$, the heavy ALP-mediated
$\mu\to 3e$ contribution has the following expression for the partial decay width:
\begin{align}
\label{eq:br-mu3e-off}
  \Gamma(\mu\rightarrow{}e a^{\ast}\rightarrow{}3e)
  =
  \frac{m_{\mu}}{128\pi^3}
  \abs{\gpe}^2
  \left(\abs{\gp}^2 + \abs{\gs}^{2}\right)
  \varphi_0(x)\,,
\end{align}
where the loop function $\varphi_0(x)$ is given in Ref.~\cite{Calibbi:2024rcm}
(also provided in Appendix~\ref{app:loop}). Another contribution to $\mu\to 3e$
arises from the interference of photonic and off-shell ALP
amplitudes~\cite{Cornella:2019uxs}. Since the ALP coupling to photon is induced
by the small coupling $\gpe$ at one loop, we neglect this interference contribution.

Another important LFV probe in the on-shell region, $2m_e < m_a < m_{\mu}-m_e$,
is the decay $\mutoegg$. The corresponding decay rate is given by an expression
analogous to Eq.~\eqref{eq:br-mu3e-on}, where the partial decay width of the ALP
into photons is
\begin{align}
  \label{eq:br-alp-photon}
  \Gamma(a \rightarrow{}\gamma\gamma)
  =
  \frac{\alpha^2 m^3_a}{16\pi^3}
  \abs{c^{\mathrm{loop}}_{\gamma\gamma}}^2,
\end{align}
with $c^{\mathrm{loop}}_{\gamma\gamma}$ defined in Eq.~\eqref{eq:c-gamgam}.

For ALP masses below $2m_e$, photons are the only SM particles into which the
ALP can decay. Since the ALP-photon coupling is loop-suppressed, the ALP width
from Eq.~\eqref{eq:br-alp-photon} is small, leading to a long ALP lifetime and
decays outside the detector. In this mass region, the search for missing energy
in the muon decay $\mutoeinv$ provides the relevant LFV probe, with the
corresponding partial width of the two-body decay $\mutoea$ given by
Eq.~\eqref{eq:br-mutoea}.

Additionally, if the ALP has interactions with a light dark sector, such as in
Eq.~\eqref{eq:ALP-DM}, and $m_a > 2 m_{\chi}$, then the ALP can decay to dark
sector particles, inducing the resonant decay $\mutoea, a\rightarrow{}\chi\chi$.
The total decay rate of $\mutoeinv$ is then calculated as
\begin{align}
  \label{eq:br-mu2einv}
  \Gamma(\mutoeinv)
  =
  \Gamma(\mu\rightarrow{}ea)\,
  \frac{1}{\Gamma_a}
  \left(\Gamma(a\rightarrow{}\chi\chi)
  + e^{{-d}/{\gamma c\tau }}\, \Gamma(a\rightarrow{}\text{visible})
  \right)
\end{align}
where in the second term visible implies $\gamma\gamma$ and, if kinematically
accessible, $ee$  as well.

\textbf{\textit{\texorpdfstring{$\boldsymbol{\Delta L=2}$}{Delta L=2} transitions.}}
ALPs with $\mu$-$e$ flavor-violating couplings also induce muonium-to-antimuonium
transitions at tree level~\cite{Endo:2020mev}. Adapting the results of
Refs.~\cite{Endo:2020mev,Bauer:2021mvw,Calibbi:2024rcm} for the $\MtoMbar$ transition
probability, we find:
\begin{align}
  \label{eq:prob-MMbar}
  \mathcal{P}({\MtoMbar})
  = \frac{8}{\pi^2 a^6_B \Gamma^2_{\mu}
  [(m^2_{\mu} - m^2_a)^2 + \Gamma^2_a m^2_a]}
  & \Bigl[
    |c_{0,0}|^2\,
    \abs*{(\gs)^2 - \delta^+_B (\gp)^2 }^2
    \notag\\
  & + |c_{1,0}|^2\,
    \abs*{(\gs)^2 - \delta^-_B (\gp)^2}^2
    \Bigr],
\end{align}
where $a_B \simeq 2.69 \times 10^5~\gev^{-1}$ is the Bohr radius of muonium and
$\Gamma_{\mu} \simeq 3 \times 10^{-19}~\gev$ is the muon decay width. The
coefficients $c_{J, m_J}$ determine the population of muonium in the angular
momentum state $(J, m_J)$. For the MACS experiment, which operated with an
external magnetic field $B=0.1~\mathrm{Tesla}$, we have $\abs{c_{0,0}}^2 = 0.32$
and $\abs{c_{1,0}}^2=0.18$. Finally, $\delta^{\pm}_B = 1 \pm (1+X^2)^{-1}$, where
the parameter $X$ is~\cite{Endo:2020mev}:
\begin{equation}
\label{eq:Mu-X}
X
=
\frac{m_B B}{a_{1s}}
\left(
  g_e + \frac{m_e}{m_{\mu} g_{\mu}}
\right),
\end{equation}
where $\mu_B = e/2m_e$ denotes the Bohr magneton, $g_e$ and $g_{\mu}$,
each approximately equal to 2, are the magnetic moment of the electron and muon,
respectively, and $a_{1s} \simeq 1.864 \times 10^{-5}~\mathrm{eV}$ is the
$1S$ hyperfine splitting of muonium. Numerically,
$X\simeq 6.24\, B/\mathrm{Tesla}$~\cite{Bauer:2021mvw}.

\textbf{\textit{Magnetic moments of charged leptons.}} ALP-lepton couplings,
both flavor-violating and flavor-conserving, generate new contributions to the
anomalous magnetic dipole moment of charged leptons, $a_{\ell} = (g_{\ell}-2)/2$.
In particular, as we will discuss in the next section, the current measurement of $a_e$ places one of the most severe
constraints on $\gpe$, which plays a key role in restricting the allowed range of
the $\muetoee$ rate in the model. Denoting the flavor-conserving and
flavor-violating contributions by subscripts ``LFC'' and ``LFV,'' respectively,
the ALP contribution to the electron magnetic dipole moment is
$\Delta a_e = (\Delta a_e)_{\text{LFC}} + (\Delta a_e)_{\text{LFV}}$, where
\begin{align}
  (\Delta a_e)_{\text{LFC}}
  &\simeq
    -\frac{(\gpe)^2}{16\pi^2}
    \left[h_1(x_e)
    +
    \frac{2\alpha}{\pi}\left(
    \log \frac{\Lambda^2}{m_e^2} - h_2(x_e)
    \right)
    B_1(\tau_e)\right],\label{eq:magmom-e-lfc}\\
  (\Delta a_e)_{\text{LFV}}
  &\simeq
    \frac{1}{16\pi^2}\frac{m_e}{m_{\mu}}\left(
    \abs{\gp}^2 -  \abs{\gs}^2
    \right)
    g_3(x_{\mu}), \label{eq:magmom-e-lfv}
\end{align}
where the second term in $(\Delta a_e)_{\text{LFC}}$ arises from the ALP-photon
coupling induced by the electron loop. In contrast, only the LFV contribution is
present for the muon magnetic moment, which reads
\begin{align}
  \label{eq:magmom-mu-lfv}
  (\Delta a_{\mu})_{\text{LFV}}
  \simeq
  \frac{1}{16\pi^2}
  \left(\abs{\gp}^2 + \abs{\gs}^2 \right)
  h_3(x_{\mu}).
\end{align}
The loop functions appearing in the above expressions for $\Delta a_{e, \mu}$
are listed in Appendix~\ref{app:loop}.

From the perspective of new physics, both $a_e$ and $a_{\mu}$ have received
significant attention due to tensions between experimental data and SM
predictions. The theoretical determination of $a_e$ depends on the precisely
measured value of the fine-structure constant, $\alpha$. At present, $\alpha$
measurements using cesium atoms~\cite{Parker:2018vye} and rubidium
atoms~\cite{Morel:2020dww} disagree at more than the $5\sigma$ level, leading to
different SM predictions for $a_e$. The SM predictions deviate from the measured
$a_e$ value in opposite directions:
$(\Delta a_e)_{\mathrm{Rb}} = (48 \pm 30) \times 10^{-14}$ and
$(\Delta a_e)_{\mathrm{Cs}} = (-88 \pm 36) \times 10^{-14}$, at level of
$1.6\,\sigma$ and $2.4\,\sigma$, respectively. On the other hand, for $a_{\mu}$,
the latest SM prediction~\cite{Aliberti:2025beg} agrees with the current
experimental world average~\cite{Muong-2:2006rrc,Muong-2:2021ojo,Muong-2:2023cdq,
Muong-2:2025xyk,Muong-2:2024hpx,Muong-2:2021vma,Muong-2:2021ovs,Muong-2:2021xzz},
yielding $\Delta a_{\mu} = (38 \pm 63) \times 10^{-11}$, in contrast to a
previous trend~\cite{Aoyama:2020ynm}. We do not attempt to explain the anomalous
$\Delta a_e$ result in this paper. ALP explanations for this anomaly can be found
in the literature; see, e.g., Refs.~\cite{Bauer:2019gfk,Endo:2020mev}. In our
numerical analysis, we conservatively require that ALP contributions to both
$\Delta a_e$ and $\Delta a_{\mu}$ agree within a $2\,\sigma$ range of the
experimental values. For $a_e$, we use the rubidium-based measurement, which is
closer to the SM value.

\textbf{\textit{Other constraints.}} Besides the LFV constraints discussed above,
there are also strong constraints on the couplings $\gsp$ and $\gpe$ from
astrophysical observations and beam-dump experiments for light ALPs.

The ALP-electron coupling $\gpe$ can facilitate the production of light ALPs inside stars through processes such as electron-nucleus scattering, $e + N \to e + N + a$, and Compton scattering, $\gamma + e \to e + a$. ALP emission introduces a new energy-loss mechanism in stars. A significant ALP-electron coupling can result in increased stellar cooling and influence stellar evolution in ways that are inconsistent with current observations, thus establishing strong bounds on ALPs~\cite{Raffelt:1990yz,Raffelt:1994ry,Raffelt:1996wa}. 

The stellar cooling bounds are most restrictive for massless ALPs but weaken for nonzero ALP masses due to Boltzmann suppression of the cooling process. Using the results from Ref.~\cite{Calibbi:2020jvd}, we find that red giant \cite{Raffelt:1994ry,Viaux:2013lha} and white dwarf \cite{MillerBertolami:2014rka} data give $\gpe \lesssim 2.2 \times 10^{-13}$ and $\lesssim 4.3 \times 10^{-13}$, respectively, for massless ALPs. These bounds rapidly weaken for ALP masses above $1-20~\textrm{keV}$ and become effectively negligible for $m_a \gtrsim 0.1~\mathrm{MeV}$. 

For very light ALPs with $m_a < 2m_e$, the EDELWEISS~III~\cite{EDELWEISS:2018tde} and GERDA\cite{GERDA:2020emj} experiments also provide strong additional bounds on $\gpe$. Their searches are based on the {\it axio-electric effect} (similar to the photoelectric effect), where ALPs with electron coupling are absorbed in the detector material (germanium), resulting in a measurable electron recoil. Assuming ALPs are produced in the Sun (for EDELWEISS~III) or that ALPs make up all galactic dark matter, the lack of any detected signal excludes values of $\gpe$ above $\order{10^{-11}}$ for ALPs lighter than $1~\mathrm{MeV}$. We also note that, in this mass regime, cosmological bounds, e.g., from the irreducible freeze-in abundance, can be significantly stronger, reaching $\gpe \sim \order{10^{-17}}$ in part of the parameter space~\cite{Langhoff:2022bij}.

For ALP masses above $1~\mathrm{MeV}$, more relevant bounds on $\gpe$ come from core-collapse supernova (SN) and beam-dump experiments. The SN bounds are derived from energy-loss arguments. ALPs coupled to electrons are produced in particle reactions such as electron-positron fusion and electron-proton bremsstrahlung inside the SN core. ALP emission contributes to SN core cooling and affects the neutrino emission rate. The energy loss at excessively rapid rate would have shortened the neutrino signal duration from SN1987A relative to observations. Using the results of Ref.~\cite{Carenza:2021pcm}, we note that for ALPs lighter than the electron, SN1987A excludes $10^{-9} \lesssim \gpe \lesssim 10^{-8}$. The SN bound extends up to $m_a \lesssim \order{100}~\mathrm{MeV}$, excluding couplings in the range $10^{-10} \lesssim \gpe \lesssim 10^{-9}$~\cite{Carenza:2021pcm}. It should be noted that SN cooling bounds depend crucially on the explosion mechanism. If the explosion is triggered by a different mechanism, such as a collapse-induced thermonuclear explosion~\cite{Bar:2019ifz}, the SN bounds on ALPs may not apply. It should be noted a recent reassessment of SN bounds finds stronger constraints than those from the traditional cooling argument, excluding roughly $10^{-12} \lesssim \gpe \lesssim 10^{-6}$ in part of the relevant parameter space~\cite{Fiorillo:2025sln}.

Finally, for ALPs with sufficiently long lifetimes, the region $m_a > 2m_e$ is also
probed by beam-dump experiments, where ALPs are produced via bremsstrahlung,
$e^- + N \to e^- + N + a$, and detected via their decay to $e^+e^-$. For
$m_a \sim 1~\mathrm{MeV}$, beam-dump
experiments~\cite{Bechis:1979kp,Konaka:1986cb,Riordan:1987aw,Bjorken:1988as,
  Bross:1989mp,Scherdin:1991xy,Bassompierre:1995kz,Alves:2017avw} exclude
$10^{-8} \lesssim \gpe \lesssim \order{10^{-3}}$. As $m_a$ increases, the range
of beam-dump constraints narrows, roughly to
$10^{-8} \lesssim \gpe \lesssim \order{10^{-6}}$ for
$m_a \sim 100~\mathrm{MeV}$~\cite{Carenza:2021pcm}.

%%%%%%%%%%%%%%%%%%%%%%%%%%%%%%%%%%%%%%%%%%%%%%%%%%%%%%%%%%%%%%%%%%%%%%%%%%%
\section{Numerical Results and Discussion}\label{sec:results}%
%%%%%%%%%%%%%%%%%%%%%%%%%%%%%%%%%%%%%%%%%%%%%%%%%%%%%%%%%%%%%%%%%%%%%%%%%%%
We now present numerical results for the allowed range of $\mathcal{B}(\muetoee)$
induced by ALP interactions. As mentioned in the Introduction, we focus on the effects
of light ALPs. Therefore, we initially consider the mass range $m_a \lesssim 10~\gev$
(for completeness, we also explore $m_a$ up to $100~\gev$ in a more general numerical
scan later).

\begin{figure}
  \centering
  \includegraphics[width=0.98\textwidth]{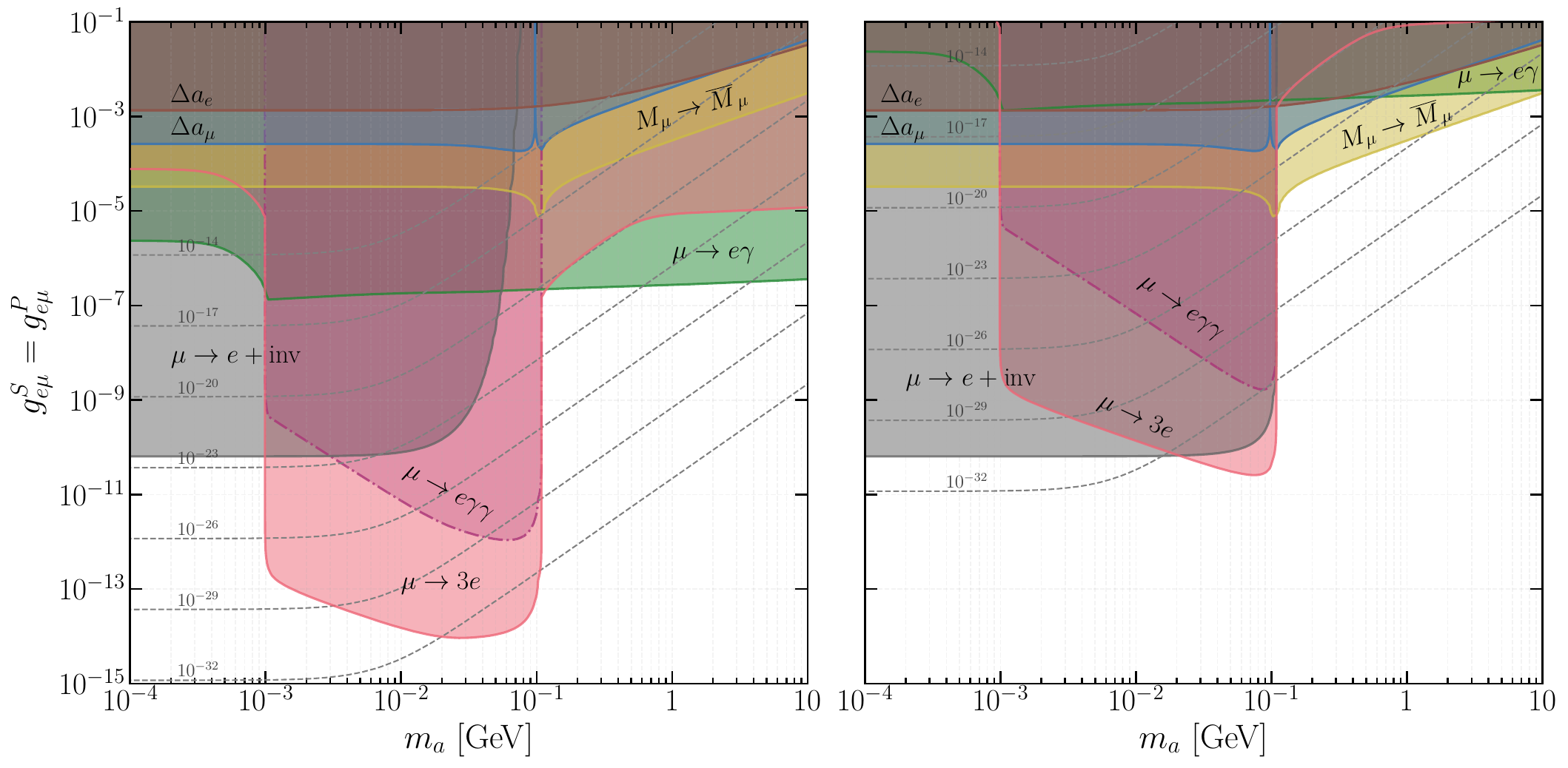}
  \caption{Constraints on the LFV couplings $\gsp$ (assuming $\gs=\gp$) as a function of
    the ALP mass. The dashed gray contours indicate $\mathcal{B}(\muetoee)$ values. The
    ALP-electron coupling is $\gpe=10^{-6}$ in the left plot and $\gpe=10^{-10}$ in the
    right plot.}
  \label{fig:constraints}
\end{figure}

In Fig.~\ref{fig:constraints}, assuming equal LFV couplings, $\gs=\gp$, we show the
leading constraints from LFV processes over a range of ALP masses for $\gpe=10^{-6}$
(left plot) and $\gpe=10^{-10}$ (right plot). The overlaid dashed gray contours show
predictions for $\mathcal{B}(\muetoee)$, ranging from $10^{-14}$ to $10^{-32}$.
Focusing first on the left plot, we see that for ALP masses below $2m_e$, the
$\mutoeinv$ search by the TWIST experiment~\cite{TWIST:2014ymv} provides the strongest
constraint (gray), excluding $\mathcal{B}(\muetoee) \gtrsim 10^{-23}$. In the on-shell
ALP mass window, $2m_e < m_a < m_{\mu} - m_e$, the ALP decays resonantly to $ee$ and
$\gamma\gamma$; the $\muetoee$ (red) and $\mutoegg$ (magenta) searches then give the
strongest bounds, ruling out $\mathcal{B}(\muetoee) \gtrsim 10^{-29}$--$10^{-32}$. For
ALP masses above the muon mass, bounds become relatively weaker: $\mutoegam$ provides
the leading constraint (green), followed by $\mueee$ and muonium oscillation (yellow).
In this region, the maximally allowed $\mathcal{B}(\muetoee)$ can approach $10^{-20}$.
The constraints from $\Delta a_e$ and $\Delta a_{\mu}$ are shown in brown and blue,
respectively. Note that $\Delta a_e$ here is sensitive to $\gpe$ only, not to the
$\gsp$ couplings, due to the assumption $\gs=\gp$ (see Eq.~\ref{eq:magmom-e-lfv}).

In the right plot of Fig.~\ref{fig:constraints}, a smaller electron coupling,
$\gpe = 10^{-10}$, is employed. The bounds from $\mueee$, $\mutoegam$, and $\mutoegg$
become comparatively weaker, as their decay rates depend on $\gpe$ (either directly or
implicitly through the electron loop-induced $a\gamma\gamma$ coupling). Likewise,
$\mathcal{B}(\muetoee)$ is also suppressed. For ALP masses below the muon mass,
constraints from $\mutoeinv$ and $\mueee$ exclude $\mathcal{B}(\muetoee) \gtrsim 10^{-31}$--$10^{-32}$. For $m_a > m_\mu$, muonium oscillation, which is insensitive to $\gpe$ and
therefore remains unchanged, becomes the leading constraint, allowing maximum
$\mathcal{B}(\muetoee) \lesssim 10^{-25}$.

Fig.~\ref{fig:constraints} illustrates the individual strengths of flavor constraints in
different ALP mass regions. It also shows that predictions for $\mathcal{B}(\muetoee)$
are suppressed, as the light ALP mass window (where the $\muetoee$ rate is enhanced due
to the light mediator) is already tightly constrained by current experimental bounds.
However, Fig.~\ref{fig:constraints} explores only a small subset of parameter space:
we assumed $\gs=\gp$ to reduce the number of free parameters and set $\gpe$ to two
specific values. To explore the ALP contributions to $\muetoee$ in more detail, we perform
a parameter scan, allowing all free parameters in the model to vary independently. We
sample $\gs, \gp, \gpe \in [10^{-18}, 10^2]$ and $m_a \in [10^{-4}~\gev, 10^2~\gev]$.
For $\gpe$, we sample values while accommodating approximate constraints in the region $m_a < m_\mu$ from  astrophysical and beam-dump searches, following the discussion in the previous section. Specifically, in the scan we exclude $\gpe \gtrsim 10^{-17}$ for $m_a < 2 m_e$, $10^{-12} \lesssim \gpe \lesssim 10^{-3}$ for $2 m_e \le m_a \le 10~\mathrm{MeV}$, and $10^{-12} \lesssim \gpe \lesssim 10^{-6}$ for $10~\mathrm{MeV} \le m_a \le m_\mu$. The lower limit of the ALP
mass is set to $10^{-4}~\gev$ to evade constraints from red giant and white dwarf data.
We vary each parameter uniformly in its allowed range to get a million
sample points, and evaluate $\mathcal{B}(\muetoee)$ and constraint observables at these points.

\begin{figure}
  \centering
  \includegraphics[width=0.99\textwidth]{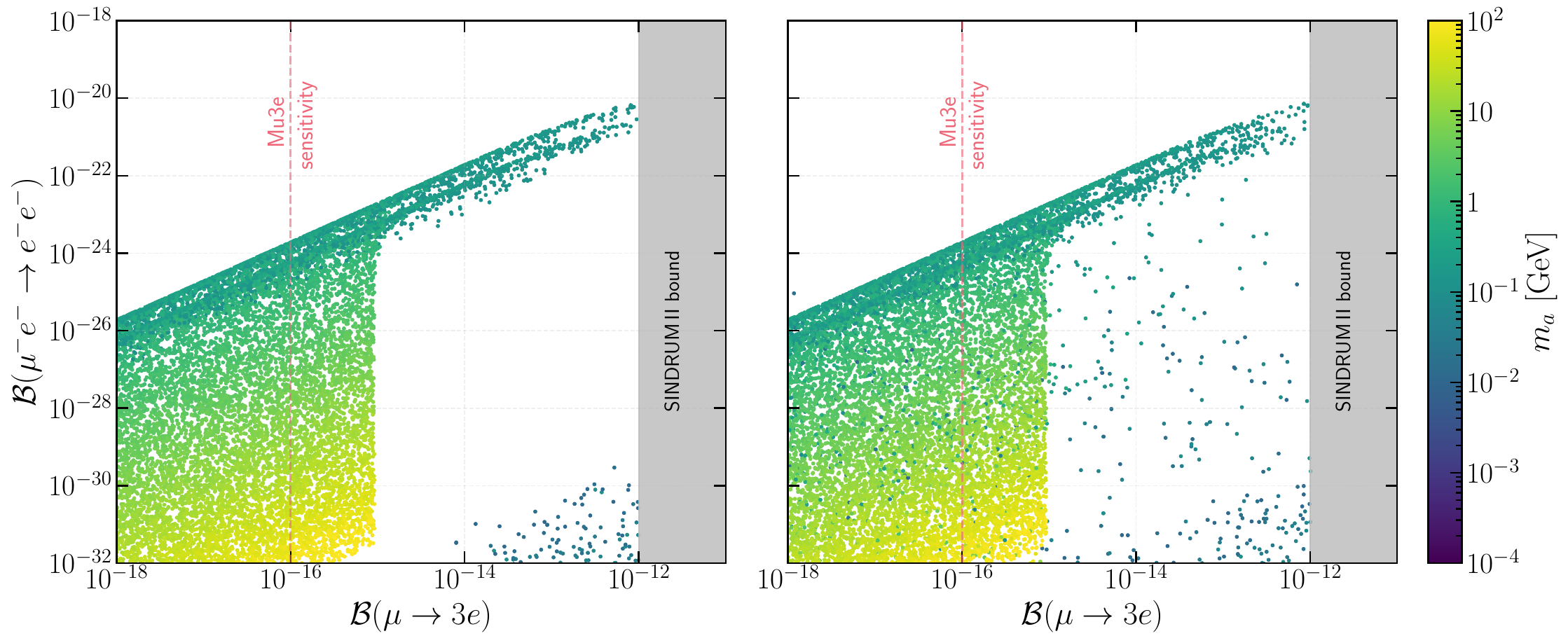}
  \caption{Scatter plot of $\mathcal{B}(\mu \to 3e)$ vs. $\mathcal{B}(\mu^- e^- \to e^- e^-)$
    for an aluminum target, with color indicating correlation with $m_a$. The coupling
    $g_{\chi\chi}$ is assumed to be zero in the left plot and finite in the right plot.
    See text for scan details.}
  \label{fig:scans}
\end{figure}
Results of this numerical scan are shown in Fig.~\ref{fig:scans} (left plot): after applying flavor constraints, we display the scatter of allowed points in the $\mathcal{B}(\muetoee)$ transition branching ratio versus $\mathcal{B}(\mueee)$ ratio, along with their correlation to the ALP mass. Since the scan over ALP couplings covers several orders of magnitude, the predicted branching ratios also span many orders, ranging from very small values to those that are unlikely to be significant for experimental searches. We present only the region corresponding to the largest $\mathcal{B}(\mueee)$ values, which could be accessible to the Mu3e experiment in the near future. The highest $\mathcal{B}(\muetoee)$ values in the scan are found to be $\lesssim 10^{-20}$, close to the current experimental limit on $\mathcal{B}(\mueee)$. Unlike the constraints shown in Fig.~\ref{fig:constraints}, we find that $\Delta a_e$ is the most effective constraint, excluding about 64\% of the sampled points mainly because of its sensitivity to $\gpe$ (see Eq.~\ref{eq:magmom-e-lfc}). The  empty region adjacent to the SINDRUM limit  largely due to the MEG~II constraint on $\mutoegam$. In Appendix~\ref{app:heatmap}, we provide more details on the impact of each constraint, along with a brief discussion of their complementarity and redundancy when combined. Also, note that in the numerical scan, we impose a stricter limit $\mathcal{B}(\mutoeinv) < 10^{-6}$  than the one from the TWIST experiment (see Table~\ref{tab:lfv}).

So far, we have considered ALPs only couple to SM particles. Let us now discuss the case where the ALPs additionally couple to a dark-sector fermion $\chi$ described by the Lagrangian in Eq.~\eqref{eq:ALP-DM}. The purpose of this additional scan is to test the robustness of our main result. An invisible ALP width suppresses the visible branching fraction $a\to e^+e^-$ and can therefore weaken the resonant $\mueee$ constraint, potentially allowing larger couplings and hence a larger $\mathcal{B}(\muetoee)$. For this parameter scan, we sample $m_a$, $\gs$, $\gp$, $\gpe$, $g_{\chi\chi}$, and $m_\chi$; the ALP couplings and ALP mass are varied in the same ranges as before. For $m_\chi$, only values for which the ALP decay $a\to \chi\chi$ is kinematically allowed are interesting, particularly when the ALP is lighter than the muon. We allow $m_\chi \in [10^{-5}\,\gev, 0.1\,\gev]$. After applying the same experimental constraints as before, the scan results are presented in Fig.~\ref{fig:scans} (right plot). From the perspective of obtaining a larger $\mathcal{B}(\muetoee)$, the results remain similar: the largest possible value for $\mathcal{B}(\muetoee)$ is again $\lesssim 10^{-20}$. It should be noted that in our analysis, the most direct constraint on $g_{\chi\chi}$ comes from the experimental limit on the branching ratio of $\mutoeinv$ [calculated using Eq.~\eqref{eq:br-mu2einv}]. A complete analysis combining various other experimental constraints\footnote{See, for example, Ref.~\cite{Darme:2020sjf} for a detailed analysis of constraints on ALPs assumed to decay primarily to invisible states.} on the ALP coupling to the dark sector, and its interplay with both LFV and flavor-diagonal ALP couplings necessary for enhanced $\muetoee$, is beyond the scope of this work and left as future work.

Before concluding, we comment on the inclusion of a diagonal muon coupling $g^P_{\mu\mu}$, defined analogously to the diagonal electron coupling in Eq.~\eqref{eq:Lag-alp}. Although it does not enter the tree-level $\mu^-e^-\to e^-e^-$ amplitude, it modifies $\mu\to e\gamma$ through the internal-muon loop and the induced $a\gamma\gamma$ coupling, and also affects $\Delta a_\mu$ and the ALP width when $a\to\mu^+\mu^-$ is kinematically allowed. We have extended the scan to include $g^P_{\mu\mu}$ with both signs. While cancellations in $\mu\to e\gamma$ occur, imposing all constraints leaves the maximal $\mathcal{B}(\muetoee)$ at the same $\mathcal{O}(10^{-20})$ level.

%%%%%%%%%%%%%%%%%%%%%%%%%%%%%%%%%%%%%%%%%%%%%%%%%%%%%%%%%%%%%%%%%%%%%%%%%%%
\section{Conclusions}\label{sec:conclusions}%
%%%%%%%%%%%%%%%%%%%%%%%%%%%%%%%%%%%%%%%%%%%%%%%%%%%%%%%%%%%%%%%%%%%%%%%%%%%
In this work, we explored the potential of the muon-conversion experiment, such as Mu2e, to investigate a nuclei-assisted charged LFV process $\mu^- e^- \to e^- e^-$ in a muonic atom within NP model with a light mediator. We employed a simplified ALP framework that includes flavor-violating couplings $g_{\mu e}^{S,P}$ and a flavor-diagonal electron coupling $g_{ee}^P$, also considering the possibility of an invisible ALP decay channel into a dark fermion pair. We derived the tree-level ALP contribution to the transition rate and provided an analytical expression for the branching ratio. As expected, the rate increases with a light mediator and scales parametrically as $(Z-1)^3$, making heavier targets inherently more favorable.

We explored the allowed parameter space of the model using relevant low-energy constraints from $\mu\to e\gamma$, $\mu\to 3e$, $\mu\to e\gamma\gamma$, $\mu\to e+\mathrm{inv}$, and muonium-antimuonium conversion, along with the anomalous magnetic moments of the electron and muon. We also considered additional astrophysical and beam-dump limits on $g_{ee}^P$ in the light-mass regime. The resulting phenomenology depends strongly on the ALP's mass. For $m_a<2m_e$, the most significant laboratory constraint is $\mu\to e+\mathrm{inv}$. In the resonant window $2m_e<m_a<m_\mu-m_e$, visible ALP decays make $\mu\to 3e$ and $\mu\to e\gamma\gamma$ especially restrictive, pushing $B(\mu^- e^- \to e^- e^-)$ to extremely small values. For $m_a>m_\mu$, the main laboratory probes are $\mu\to e\gamma$, $\mu\to 3e$, and muonium oscillations.

A key outcome of our analysis is the role of the electron's anomalous magnetic moment. Although $\Delta a_e$ may seem secondary in specific benchmark slices, the overall scan demonstrates that it is actually one of the strongest constraints on the model and the most effective veto on the sampled parameter space. This is physically clear: $\Delta a_e$ directly limits the flavor-diagonal coupling $g_{ee}^P$, which also governs the size of $\mu^- e^- \to e^- e^-$ and influences the loop-induced amplitudes relevant for other LFV observables. In this way, including $\Delta a_e$ qualitatively alters the phenomenological picture compared to a discussion based solely on LFV decay bounds.

After applying the laboratory constraints used in our scan, the maximum allowed branching ratio for $\mu^- e^- \to e^- e^-$ in aluminum drops to about $10^{-20}$, with much stronger suppression in the resonant region below the muon threshold. Therefore, while a light ALP boosts the amplitude at fixed couplings, the available parameter space is already heavily limited by existing data. We also observe a connection between $B(\mu^- e^- \to e^- e^-)$ and $B(\mu\to 3e)$: the points with the highest muonic-atom rate are close to the current $\mu\to 3e$ bound. This suggests that Mu3e will explore the most promising area of the parameter space for this process. The scan assuming a finite invisible ALP width does not significantly change this conclusion.

Overall, our results suggest that within the minimal ALP setup considered here, $\mu^- e^- \to e^- e^-$ is better seen as a complementary probe rather than the primary discovery channel. However, it remains both theoretically and experimentally well justified. It investigates the same LFV structure in a different bound-state environment, benefits from the $(Z-1)^3$ enhancement in heavy nuclei, and could offer valuable supporting evidence if a signal is first detected in another muon LFV observable. A logical next step is a target-dependent analysis using realistic atomic wave functions and experimental conditions, along with a broader examination of non-minimal ALP scenarios where additional couplings or invisible decay channels might alter the correlation pattern among LFV observables.

%%%%%%%%%%%%%%%%%%%%%%%%%%%%%%%%%%%%%%%%%%%%%%%%%%%%%%%%%%%%%%%%%%%%%%%%%%%
\section*{Acknowledgment}
This research work is supported in part by the US Department of Energy grant DE-SC0024357. We thank D.~Tedeschi for useful conversations. We also thank Edoardo Vitagliano for the helpful communication regarding  astrophysical bounds. The Feynman diagrams in this paper were drawn using \texttt{FEYNGAME}~\cite{Bundgen:2025utt}, and all plot figures were created with the help of \texttt{MATPLOTLIB}~\cite{Hunter:2007}.   

%%%%%%%%%%%%%%%%%%%%%%%%%%%%%%%%%%%%%%%%%%%%%%%%%%%%%%%%%%%%%%%%%%%%%%%%%%%

\appendix

%%%%%%%%%%%%%%%%%%%%%%%%%%%%%%%%%%%%%%%%%%%%%%%%%%%%%%%%%%%%%%%%%%%%%%%%%%%
\section{Loop Functions}\label{app:loop}
%%%%%%%%%%%%%%%%%%%%%%%%%%%%%%%%%%%%%%%%%%%%%%%%%%%%%%%%%%%%%%%%%%%%%%%%%%%
%
\begin{align}
  \label{eq:loop-fn}
  B_1(x)
  &= 1 - x f(x)^2,
    \quad\text{with }
    f(x) =
    \begin{cases}
      \arcsin\frac{1}{\sqrt{x}}
      & \text{if } x \geq 1,\\
      \frac{\pi}{2}
      + \frac{\imath}{2}\log \frac{1+\sqrt{1-x}}{1-\sqrt{1-x}}
      & \text{if } x < 1,
    \end{cases}
  \\
  g_\gamma(x)
  &= 2 \log \frac{\Lambda^2}{m_\mu^2} - 2
  - \frac{x^2 \log x}{x-1} + (x-1) \log (x-1)\,,\\
  g_2(x)
  &= 1 - 2x + 2(x-1)x \log \frac{x}{x-1}\,,\\
  g_3(x)
  &= \frac{2x^2\log x}{(x-1)^3}
    + \frac{1-3x}{(x-1)^2}\,,\\
   h_1(x)
  &= 1 + 2 x - (x-1) x\log x
    + 2x (x-3)\sqrt{\frac{x}{x-4}}\,
    \log \frac{\sqrt{x}+\sqrt{x-4}}{2}\,,\\
  h_2(x)
  &= 1 + \frac{x^2}{6}\log x - \frac{x}{3}
    - \frac{x+2}{3} \sqrt{(x-4)x}\,
    \log \frac{\sqrt{x}+\sqrt{x-4}}{2}\,,\\
  h_3(x)
  & = 2x^2 \log  \frac{x}{x-1} - 1 - 2x\,,\\
  \varphi_0(x)
  &= -\frac{11}{4} + 4x
    - \left(
    \frac{x^2}{2} \log \frac{2x-1}{x} -1 + 5x - 4x^2
    \right) \log \frac{x-1}{x} \notag\\
  & \phantom{=}\ + \frac{x^2}{2} \left[
    \operatorname{Li}_2\!\left(\frac{x-1}{2x-1}\right)
    - \operatorname{Li}_2\!\left(\frac{x}{2x-1}\right)
    \right].
\end{align}
%

%%%%%%%%%%%%%%%%%%%%%%%%%%%%%%%%%%%%%%%%%%%%%%%%%%%%%%%%%%%%%%%%%%%%%%%%%%
\section{Correlations among constraints}\label{app:heatmap}
%%%%%%%%%%%%%%%%%%%%%%%%%%%%%%%%%%%%%%%%%%%%%%%%%%%%%%%%%%%%%%%%%%%%%%%%%%
%
\begin{figure}[t]
  \centering
  \includegraphics[width=0.7\textwidth]{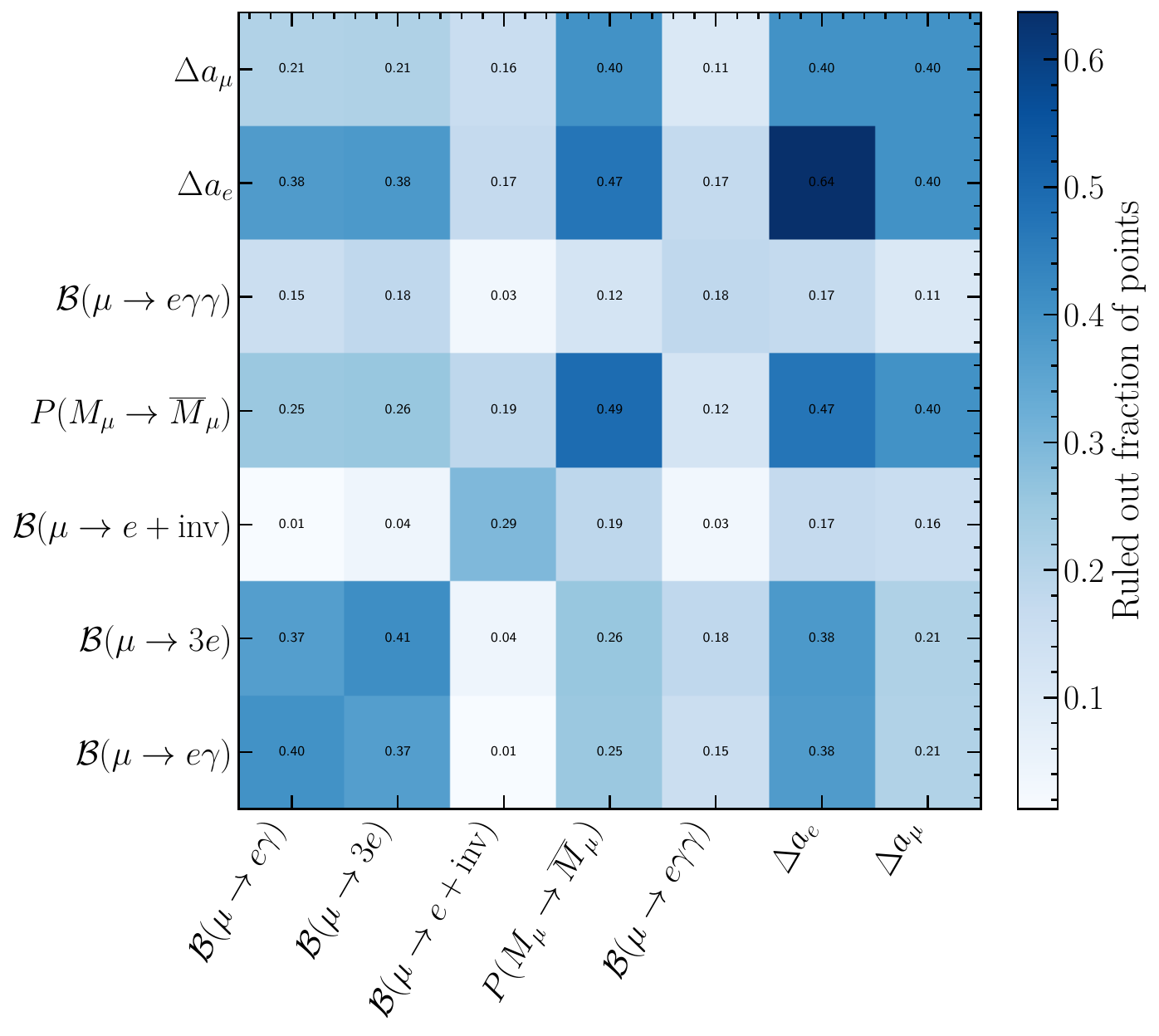}
  \caption{Heatmap of the fraction of sampled points ruled out by the constraints.
    Diagonal entries $ii$ show the fraction excluded by each constraint alone.
    Off-diagonal entries $ij$ show the fraction excluded simultaneously by both
    constraints $i$ and $j$. See text for details.}
  \label{fig:heatmap}
\end{figure}
In the numerical scan shown in Fig.~\ref{fig:scans}, several experimental constraints
were imposed to identify the allowed parameter space. In this Appendix, we quantify the
strength of each constraint and describe their complementarity or redundancy when
combined. This is visualized in the heatmap in Fig.~\ref{fig:heatmap}, which shows the
fraction of points excluded by each constraint.

The diagonal entries give the fraction of points ruled out by each constraint alone. The
largest value in the diagonal, $0.64$, corresponds to $\Delta a_e$, indicating that it
is the strongest constraint, excluding $64\%$ of the scanned points by itself. This is
followed by $\MtoMbar$ and $\mueee$, which exclude $49\%$ and $41\%$ of the points,
respectively.

Off-diagonal entries $(i,j)$ represent the fraction of points ruled out simultaneously
by constraints $i$ and $j$. A large off-diagonal entry indicates that the two constraints
target the same region of parameter space; one may therefore be redundant. For example,
$\Delta a_\mu$ excludes $40\%$ of points alone, while $\Delta a_e$ excludes $64\%$. However,
$\Delta a_e$ and $\Delta a_\mu$ jointly exclude only $40\%$ of points. This means that
including $\Delta a_e$ renders $\Delta a_\mu$ redundant: all points excluded by
$\Delta a_\mu$ are already excluded by the stronger $\Delta a_e$ constraint.

On the other hand, a small off-diagonal entry can indicate two possibilities. First, trivially,
one or both constraints may be weak (with correspondingly small diagonal values). The
other possibility is that both constraints are strong (large diagonal values) but target
different regions of parameter space, making them complementary. For example, $\mueee$
and $\mutoeinv$ exclude $41\%$ and $29\%$ of points individually, but jointly exclude
 merely $4\%$ of the \textit{same} points, revealing their complementarity.

%%%%%%%%%%%%%%%%%%%%%%%%%%%%%%%%%%%%%%%%%%%%%%%%%%%%%%%%%%%%%%%%%%%%%%%%%%

\bibliography{refs}{}

\end{document}